\documentclass{ws-ijgmmp}

\def\nn{\nonumber}

\def\be{\begin{equation}}
\def\ee{\end{equation}}
\def\bea{\begin{eqnarray}}
\def\eea{\end{eqnarray}}

\def\cL{{\cal L}}

\newcommand{\w}[1]{\\[0.#1cm]}

\begin{document}

\markboth{R. Percacci}
{Towards Metric-Affine Quantum Gravity}

%
\catchline{}{}{}{}{}
%

\title{Towards Metric-Affine Quantum Gravity}

\author{Roberto Percacci}

\address{SISSA, via Bonomea 265\\
34136 Trieste, Italy\\
and INFN, Sezione di Trieste\\
\email{percacci@sissa.it} }

\maketitle

\begin{history}
\received{(Day Month Year)}
\revised{(Day Month Year)}
\end{history}

\begin{abstract}
I review here some motivations to consider a theory
of gravity based on independent metric and connection,
and its status as a quantum theory.
\end{abstract}


\section{Why Metric-Affine Gravity?}	

Metric-Affine Gravity (henceforth MAG) is a class of theories
of gravity with independent metric and connection.
There are phenomenological motivations to study
modified theories of gravity like MAG,
\footnote{See \cite{hehl} for a comprehensive overview.}
but here we shall
only be concerned with the relation between gravity and
the other interactions.
Unlike Einstein's General Relativity (GR),
in the case of the electroweak and strong forces, 
the mediator of the interaction is a connection.
This is a good motivation for alternative theories 
of gravity where also the gravitational connection is dynamical.

GR can be formulated in many inequivalent ways.
One choice is the set of fields one works with.
Unimodular metric, metric, tetrad are just some of the possibilities,
with increasing number of fields.
When one increases the number of fields, however, one also enlarges 
the gauge group, in such a way that the number of physical degrees of freedom remains always the same.
Thus, these different formulations are all equivalent,
independently of the form of the action:
they just amount to different partial gauge fixings of the same theory.
\footnote{Alternatively, one can think of them as the result
of applying the St\"uckelberg trick to unimodular gravity,
which is the formulation with the least number of fields
that is still local \cite{Gielen:2018pvk}.}

Another way to increase the number of dynamical fields,
without actually changing the physics, 
is to allow the connection to be independent of the metric, or tetrad,
as in Palatini and Einstein-Cartan formulations of GR.
\footnote{For simplicity we consider here pure gravity
without matter. } 
In these cases the elimination of the additional fields is the result of the equations of motion.
This, however, is a peculiarity of theories with a Lagrangian
that is linear in curvature.
When we allow Lagrangians quadratic in curvature,
the theory is no longer equivalent to a purely metric theory.
We are then in the realm of MAGs in the proper sense.

Just as gauge theories can be in different phases,
also gravity can, although our understanding is much more
limited in this case.
We shall see that MAGs look like
gauge theories in a Higgs phase.
This provides a partial answer to the question
``what is the origin of the Planck scale'',
that is analogous to the answer that the Standard Model (SM) gives
to the question ``what is the origin of the Fermi scale''.
At the core of our arguments lies a dynamical understanding of the
conditions of metricity and torsionlessness, that in GR are
merely postulated.

\section{Gauge theories in the Higgs phase}

We start by considering the
paradigmatic example of a superconductor.
It is well-known that the macroscopic properties of
a superconductor follow simply from the assumption that
the electromagnetic $U(1)$ gauge group is in the Higgs phase
\cite{weinberg}.
There is a complex scalar field $\phi=\rho e^{i\varphi}$ with nonzero VEV.
The radial mode $\rho$ has a mass and can be ignored at sufficiently
low energy.
Infrared physics only depends on the phase $\varphi$, which behaves like a Goldstone boson.
It transforms under $U(1)$ as 
$\varphi(x)\to\varphi(x)+\alpha(x)$
and has a covariant derivative 
$D_\mu\varphi=\partial_\mu\varphi-A_\mu$.
It is natural to assume that the Lagrangian for
$\varphi$ contains a term quaratic in $D\varphi$,
so, in a static situation, its energy will be minimized
if the vacuum is such that
\be
D_i\varphi=0\ .
\ee
This is the main property of the superconducting state.
It implies that the magnetic field vanishes in the
bulk of the material. This is known as the Meissner effect.
The vanishing of resistivity can also be derived from this property.
If one chooses the gauge so that $\varphi$ is constant,
the kinetic term of $\varphi$ is seen to become a mass
for the longitudinal part of the photon: this is the essence
of the Higgs phenomenon.

A non-abelian version of this occurs in the Weinberg-Salam electroweak theory, and more generally whenever a gauge theory with
gauge group $G$ is in the Higgs phase, with unbroken subgroup $H$
\cite{appelquist,longhitano}.
There is an order parameter, a scalar field $\phi$ in some
representation of $G$, and a potential that is minimized
when $\phi$ lies in an orbit of $G$ that is diffeomorphic
to $G/H$.
The potential gives a mass to the radial degrees of freedom of $\phi$.
At sufficiently low energy these massive degrees of freedom
can be ignored and one only has the fields $\varphi^\alpha$
with values in the coset space $G/H$. These are the Goldstone bosons.
Their dynamics is given by the Lagrangian
\be
\cL_g=-\frac{f^2}{2}D_\mu\varphi^\alpha
D^\mu\varphi^\beta h_{\alpha\beta} 
\qquad\mathrm{where}\mathrm
\qquad D\varphi=\partial\varphi^\alpha+gA^a K_a^\alpha(\varphi)\ ,
\ee
and $h_{\alpha\beta}$ is an invariant metric in $G/H$.
The Lie algebra of $G$ is the direct sum of the Lie
algebra of $H$ and a space $\mathcal{P}$.
Thus we can decompose the Yang-Mills field 
$A=A|_{\mathcal{L}(H)}+A|_{\mathcal{P}}$.
In the unitary gauge where $\varphi=\varphi_0$ is constant,
\be
D\varphi_0^\alpha=g A^a|_{\mathcal{P}}K_a^\alpha (\varphi_0)
\label{eleonora}
\ee
and the kinetic term of the Goldstone bosons becomes
a mass term for the $\mathcal{P}$-component of the gauge field:
\be
\cL_g=-\frac{1}{2}m_A^2 \sum_{a\in{\cal P}}A_\mu^a A^{\mu a}\ ,
\quad\text{where}\ m_A=gf\ .
\ee
We call this a ``Higgsless Higgs mechanism'',
because there is no Higgs field left over.
As in superconductivity, the basic low energy property 
of the vacuum
is that the Goldstone bosons are covariantly constant.
This is a gauge-invariant statement; in the unitary gauge
it becomes the statement that the $\mathcal{P}$-components 
of the gauge field vanish. 
Only the $\mathcal{L}(H)$ gauge field remains
massless and is dynamical.

The standard Higgs mechanism is obtained by going back to
the field $\phi$ carrying a linear representation of $G$
({\it i.e.}, in geometrical language, isometrically embedding
$G/H$ in a vectorspace).
For example, if the potential is 
$V=\frac{\lambda}{4}(\rho^2-f^2)^2$ 
with $\rho=|\phi|$,
the Higgs scalar $\rho$ has a mass $m_\rho=\sqrt\lambda f$
and the Higgsless Higgs mechanism is a good description of
physics at energy $E\ll m_\rho$.
In addition, at energies $E\ll m_A$, one can set to zero
the $\mathcal{P}$-components of the gauge field.
This is a property of the electroweak vacuum at energies
below approximately 100 GeV.	
\footnote{These statements about local physics
have well-known geometrical counterparts in the the following theorems:
a principal bundle $P$ has an $H$ structure iff there exists a global section $\sigma$ of the associated bundle $P\times_G G/H$.
Furthermore, a connection $A$ in $P$ is an $H$
connection iff $\sigma$ is covariantly constant. See \cite{KN}.}

\section{The Higgs phenomenon in MAG}

\subsection{Fields and gauges}

Before introducing a metric,
parallel transport in gravity is given by a linear connection,
so, a priori, the analog of the gauge group $G$ for gravity is
the linear group $GL(4)$.
However, in the standard metric and tetrad formalisms,
invariance under $GL(4)$ is not manifest.
This is because one usually works with special classes
of linear bases in the tangent bundle:
either coordinate bases $\{\partial_\mu\}$
or orthonormal bases (tetrads) $\{e_a\}$.
For our purposes it is convenient to have manifest $GL(4)$
invariance \cite{Floreanini:1989hq,higgs}, 
so we shall work with generic linear bases,
or ``frames'', $\{\theta_a\}$, 
where $\theta_a=\theta_a{}^\mu\partial_\mu$,
or equivalently the dual co-frames $\theta^a=\theta^a{}_\mu dx^\mu$
that shall also be called the ``soldering form''.
The frame (or co-frame) field will be a dynamical variable, generalizing the tetrad.
In additon we have a dynamical metric, whose components in the
frame $\{\theta_a\}$ are $\gamma_{ab}$.
Finally, there is a dynamical connection with components
$A_\mu{}^a{}_b=\theta^c{}_\mu A_c{}^a{}_b$.

Let us point out right away that, already at the kinematical 
level, the frame and the metric
are non-linear variables, because of the constraints:
\bea
&&\bullet\  \text{the metric}\ \gamma_{ab}\ \text{has signature}\ -,+,+,+\ ;
\\
&&\bullet\ \text{the frame field}\ \theta^a{}_\mu 
\text{is non-degenerate}:\  
\det\theta\not=0\ .
\label{constraints}
\eea
In particular, also $\gamma$ is non-degenerate.
This means that, locally, the metric is a field with values
in the coset space $GL(4)/O(1,3)$
and the frame has values in $GL(4)$.
For this reason we will refer to these fields as gravitational Goldstone bosons.
\footnote{One could try to see them as the result of breaking $GL(4)$ in a model with global linear invariance, but this interpretation is not so useful. To justify the terminology, it is enough that the
gauge group acts transitively on these variables,
as we see in (\ref{glsol},\ref{glmet}).}

The components of the metric and connection in coordinate basis
are given by
\bea
g_{\mu\nu}&=&\theta^a{}_\mu\, \theta^b{}_\nu\, \gamma_{ab}\ ,
\nn\\
A_\lambda{}^\mu{}_\nu&=&\theta^{-1}{}_a{}^\mu A_\lambda{}^a{}_b \theta^b{}_\nu
+\theta^{-1}_a{}^\mu \partial_\lambda \theta^a{}_\nu\ .
\label{composites}
\eea
They can be viewed as composite fields of the fundamental variables
$\theta$, $\gamma$ and $A$.

The gauge group of the theory consists of automorphisms 
of the bundle of linear frames. Locally, they are given by diffeomorphisms and local changes of frame.
The latter act on the fields as follows:
\bea
A'_\mu{}^a{}_b&=&\Lambda^{-1a}{}_c A_\mu{}^c{}_d\Lambda^d{}_b
+\Lambda^{-1a}{}_c \partial_\mu\Lambda^c{}_b\ ,
\\
\theta^{\prime a}{}_\mu&=&
\Lambda^{-1a}{}_c\theta^c{}_\mu\ ,
\label{glsol}
\\
\gamma'_{ab}&=&\gamma_{cd}\Lambda^c{}_a\Lambda^d{}_b\ .
\label{glmet}
\eea
In particular, choosing $\Lambda=\theta$, 
we see that we can enforce the gauge condition
$\theta^a_\mu=\delta^a_\mu$.
This brings us to a coordinate basis and completely breaks $GL(4)$.
In this gauge (\ref{composites}) shows that there is no
difference between latin and greek indices.
This corresponds to the standard formulation of gravity
in terms of a metric, possibly with an independent connection.
This will be called the metric gauge.

Alternatively one can choose the gauge so that
$\gamma_{ab}=\eta_{ab}=\mathrm{diag}(-1,1,1,1)$,
leaving an unbroken $O(1,3)$ gauge group.
This means that we are using an orthonormal frame.
Equations (\ref{composites}) are the usual relations
holding in the tetrad formalism, relating the components
of the metric and connection in a coordinate frame to those
in an orthonormal frame.
This will be called the tetrad gauge.

It is crucial that there is not enough gauge freedom to fix both gauges simultaneously.
Therefore, unlike the theories of Section 2, 
where the Goldstone bosons
were completely absorbed in the longitudinal part of the gauge field,
one of the two Goldstone bosons remains
as a physical dynamical variable at low energy.

\subsection{Lagrangian and VEVs}

We can now introduce torsion and non-metricity
as the covariant derivatives of the Goldstone bosons:
\bea
\Theta_\mu{}^a{}_\nu&=&
\partial_\mu \theta^a{}_\nu-\partial_\nu \theta^a{}_\mu+
A_\mu{}^a{}_b\, \theta^b{}_\nu
-A_\nu{}^a{}_b\, \theta^b{}_\mu\ ,
\\
Q_{\lambda ab}&=&
-\partial_\lambda \gamma_{ab}
+A_\lambda{}^c{}_a\, \gamma_{cb}
+A_\lambda{}^c{}_b\, \gamma_{ac}\ .
\eea

Note that the meaning of these tensors is obscured
if one works with coordinate or orthonormal frames.
Indeed, in the metric gauge (coordinate frames)
torsion appears to be just an algebraic combination
of connection components
\be
\Theta_\mu{}^\rho{}_\nu=A_\mu{}^\rho{}_\nu
-A_\nu{}^\rho{}_\mu\ ,
\ee
whereas in the tetrad gauge (orthonormal frames)
it is the non-metricity that appears to be a purely algebraic
combination of connection components:
\be
Q_{\lambda ab}=A_{\lambda ab}+A_{\lambda ba}\ .
\ee
In order to recognize that these tensors play very similar roles,
it is necessary to work with generic frames.

Let us now come to the dynamics.
The Goldstone bosons can be viewed as a special type of matter,
and as in any gauge theory coupled to matter,
the Lagrangian has terms quadratic in curvature
and terms quadratic in the covariant derivatives of the
matter fields.
Thus the action of MAG is
$$
S_{MAG}=\int d^4x\sqrt{-g}\,\cL_{MAG}\ ,
\qquad
\cL_{MAG}=\cL_P+\cL_{F}+\cL_{TQ}\ .
$$
Here $\cL_{P}=-m_P^2F_{ab}{}^{ab}$ 
is the Palatini Lagrangian,
\bea
\cL_{F}&=&
F^{\mu\nu\rho\sigma} \big( c_1 F_{\mu\nu\rho\sigma} 
+ c_2 F_{\mu\nu\sigma\rho} 
+ c_3 F_{\rho\sigma\mu\nu} 
+ c_4 F_{\mu\rho\nu\sigma} 
+ c_5 F_{\mu\sigma\nu\rho} 
+ c_6 F_{\mu\sigma\rho\nu} \big)
\nn\w2
&&  \!\!\!\!
+ L^{\mu\nu} \big(c_7 L_{\mu\nu} + c_8 L_{\nu\mu} \big)
+ K^{\mu\nu} \big( c_9 K_{\mu\nu} 
+ c_{10} K_{\nu\mu}\big) 
+ K^{\mu\nu}\big(c_{11} L_{\mu\nu}
+ c_{12} L_{\nu\mu} \big)
\nn\w2
&& \!\!\!\!
+F^{\mu\nu}\big(c_{13} F_{\mu\nu}
+ c_{14} L_{\mu\nu}
+ c_{15} K_{\mu\nu}\big)
+c_{16}L^2\ ,
\eea
where
$F_{\mu\nu}=F_{\mu\nu\lambda}{}^\lambda$,
$K_{\mu\nu}=F_{\lambda\mu\nu}{}^\lambda$,
$L_{\mu\nu}=F_{\lambda\mu}{}^\lambda{}_\nu$,
$L=L^\mu{}_\mu=-K^\mu{}_\mu$
and
\bea
\cL_{TQ}&=& T^{\mu\rho\nu} \big(a_1 T_{\mu\rho\nu} + a_2 T_{\mu\nu\rho}\big)  + a_3 T^\mu T_\mu
+Q^{\rho\mu\nu}\big( a_4 Q_{\rho\mu\nu} 
+ a_5 Q_{\nu\mu\rho}\big)  
\w2
&& \!\!\!\!
+ a_6 Q^\mu Q_\mu   
+ a_7\tilde Q^\mu \tilde Q_\mu 
+  a_8 Q^\mu \tilde Q_\mu  
 + a_9 T^{\mu\rho\nu} Q_{\mu\rho\nu} 
+ a_{10} T^\mu Q_\mu 
+ a_{11} T^\mu \tilde Q_\mu\ ,
\nonumber
\eea
where
$T_\mu=T_\lambda{}^\lambda{}_\mu$,
$Q_\mu=Q_{\mu\lambda}{}^\lambda$,
$\tilde Q_\mu=Q_\lambda{}^\lambda{}_\mu$.

Here $\cL_F$ is a generalization of the Yang-Mills Lagrangian and
$\cL_{TQ}$ are the kinetic terms of the Goldstone bosons.
The Palatini term has no analog and diffeomorphism invariance
forbids a potential term for $\gamma$ and $\theta$, except
for a cosmological term, that we shall ignore.

As a ``vacuum'' we choose flat Minkowski space:
$F_{abcd}=0$, $T_{abc}=0$, $Q_{abc}=0$.
In a suitable gauge, it can be represented as
$A_\mu{}^a{}_b=0$, $\gamma_{ab}=\eta_{ab}$, 
$\theta^a{}_\mu=\delta^a_\mu$.
Denoting $a_\mu{}^a{}_b$ the fluctuation of the connection, 
we see that
\be
\Theta_\mu{}^a{}_\nu=\ 
a_\mu{}^a{}_\nu-a_\nu{}^a{}_\mu\ ,\qquad
Q_{\mu ab}=\ 
a_{\mu ab}+a_{\mu ba}\ .
\ee
These equations are analogous to (\ref{eleonora}).
Then $\cL_{TQ}$ just becomes a mass term for the connection.
The curvature also contains a term quadratic
in the gauge field, so that the Palatini lagrangian
also contributes to the mass term
$-m_P^2(a_a{}^{ae}a_{be}{}^b-a_{bae}a^{aeb})$.
It is natural to assume that all the coefficients
$a_1,\ldots a_{11}$ are of the order of $m_P^2$.
For generic values of the coefficients the mass matrix
will be non-degenerate and all components of the connection
will become massive, with a mass of order of the Planck mass.
This is a gravitational version of the ``Higgsless'' Higgs phenomenon,
because it involves only Goldstone bosons and no ``Higgs'' particle
\cite{higgs,kirsch,leclerc}.

\subsection{The low energy effective field theory}

The Higgs phenomenon removes the connection degrees of freedom
from the spectrum below the Planck mass.
However, we have already observed that it is not possible to
remove both Goldstone bosons by a choice of gauge.
One of them remains as a low energy massless degree of freedom.
Let us now consider the low energy dynamics of this Goldstone boson.

Here we recall that given $\theta$, $\gamma$, there is a unique 
connection $\mit\Gamma$ 
such that $\bar\Theta=0$, $\bar Q=0$.
It is called the Levi--Civita Connection and its components
in a general frame are
\be
\mit\Gamma_{abc}={1\over2}\bigl(
E_{cab}+E_{abc}-E_{bac}\bigr)
+{1\over2}\bigl(C_{abc}+C_{bac}-C_{cab}\bigr)\ ,
\ee
where
$E_{cab}={\theta^{-1}}_c{}^\lambda\, \partial_\lambda\gamma_{ab}$
and
$C_{abc}=\gamma_{ad}\, \theta^d{}_\lambda\bigl({\theta^{-1}}_b{}^\mu\,
\partial_\mu{\theta^{-1}}_c{}^\lambda-
{\theta^{-1}}_c{}^\mu\, \partial_\mu{\theta^{-1}}_b{}^\lambda\bigr)$.
In a coordinate frame $C_{abc}=0$ and this formula reduces to
the Christoffel symbols, while in an orthonormal frame
$E_{abc}=0$.
Any connection $A$ can be split uniquely in
$A={\mit\Gamma}+\Phi$, where $\Phi$ is a tensor.
Then, in the action we may replace $A$ 
by $\Phi$.
Thus we can write 
$S(A,\theta,\gamma)=S({\mit\Gamma}(\theta,\gamma)+\Phi,\theta,\gamma)=S'(\Phi,\theta,\gamma)$.
Now we note that
\be
\Theta_\mu{}^a{}_\nu=\ 
\Phi_\mu{}^a{}_\nu-\Phi_\nu{}^a{}_\mu
\ ,\qquad
Q_{\mu ab}=\ 
\Phi_{\mu ab}+\Phi_{\mu ba}
\ee
and therefore, without making any assumptions about the
nature of the vacuum, $\cL_{TQ}$ is seen to be a mass term for $\Phi$.
The mass matrix is the same as the mass matrix for $a$
when we expanded about the flat space. We assume that
it is generic and therefore nondegenerate.
This implies that, for any $\theta$ and $\gamma$,
the deviation of $A$ from the Levi-Civita connection
is massive, with a mass that one may reasonably presume
to be of the order of the Planck mass.
\footnote{This is because the Palatini term already
gives a Planck mass to some components of $A$.
It would be highly interesting is some components of $A$
had a mass that is much below the Planck scale, but we
shall not discuss this possibility here.}
For all sub-Planckian physics it is therefore
a very good approximation to assume that $\Phi=0$,
or in other words $\Theta=0$ and $Q=0$.
In this way we understand that these conditions,
which in GR are simply postulated, are natural properties of MAG
at low energy.

If we impose that $\Theta=0$ and $Q=0$, the curvature
tensor $F$ reduces to the Riemann tensor constructed with
$\theta$ and $\gamma$ (or equivalently with the metric $g$),
and the Lagrangian $\cL_F$ reduces to the Lagrangian
of Higher Derivative Gravity (HDG),
whereas $\cL_P$ reduces to the Hilbert Lagrangian.
In this context, the explanation for the stiffness of
spacetime - its resistance to being curved - is due to the
large value of the Planck density, compared to ordinary
energy density.
Indeed, at low energy, the HDG action is negligible with
respect to the Hilbert action and the equations of motion
say that ``curvature\,$\approx m_P^{-2} T$'',
where $T$ is the energy density of matter.
It takes an energy density of order of the Planck density
to produce a curvature of the order of $m_P^2$.
From the point of view of MAG, the stiffness of spacetime
is very reminiscent of the Meissner effect.
The vacuum has no curvature and due to the mass of the gauge field,
a point-like disturbance will generate a curvature
that decays exponentially within a Planck length.

MAG is much more similar to the theories of
the electroweak and strong interaction
than GR, and this somewhat shifts the focus of the questions
that one would like to be answered by quantum gravity.
In QCD, crucial issues are confinement and chiral symmetry breaking.
Even though a detailed explanation is lacking,
they are believed to be due to the complicated, 
strong dynamics at low energy.
In particular, the strong scale emerges from the phenomenon of ``dimensional transmutation''.
\footnote{See \cite{holdom} for possible
analogies between QCD and gravity.}
The central issue of the electroweak theory is the
origin of ``electroweak symmetry breaking''.
As long as the Higgs particle had not been discovered,
one could simply have modelled it by the Higgsless theory
described in Section 2, where the order parameter carries 
a nonlinear realization of the gauge group.
This theory has the drawback of not being renormalizable.
Even worse, it becomes strongly coupled and breaks down
at energies comparable to the Higgs VEV.
In the SM, the order parameter carries a linear realization
and the theory is renormalizable. 
For the time being, this description in terms of an elementary
scalar field is completely satisfactory,
but this may not remain the case indefinitely.
In fact, we have the example of superconductivity,
where the Landau-Ginzburg description in terms of
a complex order parameter is known to be only an effective
low energy approximation of BCS theory, where the scalar
is viewed as a condensate of pairs of electrons.

Since the order parameter for the gravitational Higgs phenomenon
is the metric, the central question for a
quantum theory of spacetime is:
why is the metric nondegenerate,
and more precisely, why does it have signature $-+++$?
\cite{Floreanini:1990se}.
This question cannot be answered within
the present formulation of MAG,
which is based on nonlinear (Goldstone) fields.
One may begin to answer it by extending the validity of the
theory towards higher energies.
In Landau-Ginzburg theory and in the SM, a more fundamental
description is given in terms of the linearly transforming
field $\phi$, and going from $\varphi$ to $\phi$
one has to enlarge the number of dynamical fields.
In the case of gravity, this is not necessary,
because the constraints (\ref{constraints}) are formulated
as inequalities rather than equalities.
To go from the nonlinearly realized theory to the linear one,
in gravity, it is enough to relax those constraints.
No new degrees of freedom are needed.
Unfortunately, when one relaxes these constraints,
the metric may become degenerate or even zero,
and it is very difficult to formulate dynamics
under these circumstances.
Some attempts to derive the constraints (\ref{constraints})
from a self-consistent bi-metric dynamics were made in 
\cite{mfqg,flop2}.

\section{Towards Quantum MAG}

Let us discuss a little more the properties of MAG.
In order to minimize the number of independent variables,
we can work in the metric gauge (i.e. with coordinate frames),
where the dynamical variables are the metric $g_{\mu\nu}$
and the independent linear connection $A_\lambda{}^\mu{}_\nu$.
When linearized around flat space, it is found to propagate
many different degrees of freedom, depending of the values of
the coefficients $c_i$ and $a_i$.
In general, the metric fluctuation $h_{\mu\nu}$
contains one $2^+$, one $1^-$ and two $0^+$ states
(where $J^P$ denotes a state with spin $J$ and parity $P$).
The fluctuation of the connection $a_{\lambda\mu\nu}$
contains one $3^-$ state, three $2^+$, two $2^-$,
six $1^-$, three $1^+$, four $0^+$ and one $0^-$.
In contrast to the Yang-Mills Lagrangian, 
that uses a positive definite inner product in the Lie algebra, 
the Lagrangian $\cL_{MAG}$ contains in general ghosts and
tachyons.
(This is what also happens, but for different reasons, in HDG).
It is therefore interesting to find ranges of values
of coefficients for which there are no ghosts and tachyons.
Given that the general MAG action depends on 28 parameters,
this is an extremely complicated issue.
It has been partly solved in two special cases,
namely when one imposes $Q=0$ 
\cite{svn,lhl} or $T=0$
\cite{persez}.
Using $F=R+\nabla\Phi+\Phi^2$ we can rewrite
$S(g,A)=S'(g,\Phi)$
where, schematically,
$$
S'=
\!\!\!\int d^4x\sqrt{|g|}\big[R+\Phi^2+R^2+R\nabla\phi+R\Phi^2
+(\nabla\Phi)^2+\nabla\Phi\Phi^2+\Phi^4\big]\ .
$$
It is interesting to observe that when one rewrites the
action in this way, the ghost- and tachyon-free
actions turn out not to contain any terms $\sim R^2$.
Thus, the origin of the ghosts and tachyons in MAG and in HDG
are, at least in some cases, related.
Unfortunately, there appears to be no reason for
these special conditions to be radiatively stable,
so that the issue of the ghosts and tachyons remains open.
We will discuss this further below.

Classical scale invariance is achieved by putting
$m_P=a_1=a_2=\ldots =a_{11}=0$.
This leaves the free parameters $c_1$, $c_2$ \ldots $c_{16}$.
In fact, if we postulate $\delta A_\mu{}^\rho{}_\nu=0$,
as in Yang-Mills theory, $\cL_F$ is even Weyl-invariant.
Scale invariance is in any case broken in the quantum theory because 
of the running of couplings under the Renormalization Group (RG).
However, there may be Fixed Points (FP),
where the beta functions vanish and
one achieves quantum scale invariance
\cite{wetterich,morris}.
If a RG trajectory reaches a FP in the UV, it describes
a UV complete theory.
There can be free fixed points, leading to asymptotic freedom (AF)
and interacting fixed points, leading to asymptotic safety (AS).

Due to the technical complication, no systematic analysis of
the RG of MAG has been undertaken so far.
The literature on AS of gravity focuses almost entirely on metric theories.
At one loop HDG can be AF (depending on the sign of some parameter) \cite{avrabar}.
This result has been reproduced using the Functional RG,
but in this case Newton's coupling is nonzero at the FP
\cite{codello,niedermaier,op}.
Whether the coefficients of $R^2$ are AF or AS in a more
sophisticated approximation, is a question
that is currently still being studied, but the expectation is
that there should exist a FP where all couplings are nonzero
\cite{bms}.
There have been some calculations of beta functions with the
Functional RG in a theory
with independent connection \cite{bs,dr1,dr2,hr,pp}
but none in the full MAG with running couplings $c_i$.
Altogether it seems that quantum scale invariance in MAG
is possible and even likely, but much more work is needed to
establish this.

If AS can be achieved in MAG, we remain with the issue of ghosts,
and more generally of unitarity.
Over time, there have been many proposals on how to circumvent
the ghost problem,
see \cite{Mannheim:2011ds,anselmi,Donoghue:2019fcb}
for some recent proposals.
In the literature on AS gravity, where the connection is
always constrained to be the Levi-Civita connection,
but higher derivative terms are present,
one common point of view is that ghosts are artifacts 
of a finite truncation: if one could keep all terms in
the action, the propagator would be an entire function with
only one pole at zero.
Another possibility is that ghosts are an artifact of the expansion around flat space \cite{br}.
In the context of MAG, where the Planck scale appears as the
mass of the connection, the following mechanism may be at work:
the physical (pole) mass of a state is defined as the
running mass evaluated at the pole mass itself:
$m_{\mathrm{phys}}=m(k=m_{\mathrm{phys}})$.
If the theory achieves scale invariance at high energy,
the running mass runs like the cutoff $k$ itself,
and this implicit equation may well not have any solution \cite{flop3}.
We note that this mechanism is actually blind to the
sign of the residue, so that it may eliminate both ghosts
and healthy states.
In fact it suggests that no particle could exists with mass
comparable or higher than the Planck mass.
Interestingly, a similar conclusion has been reached  
also in a context of non-commutative geometry \cite{lizzi}.
If scale invariance at high energy removes all heavy particles 
from the physical spectrum, the whole phenomenology of MAG
could be very different from what is normally envisaged
\cite{Percacci:2004yy}.

Lastly, it is worth recalling that MAG can be easily generalized
to construct unified theories of gravity and Yang-Mills interactions
\cite{higgs,Nesti:2007ka,Nesti:2009kk}.
To this end, let us consider the subclass of MAGs with $Q=0$.
Then, the gauge group of the theory is the Lorentz group $SO(1,3)$.
All other interactions can be unified in the orthogonal
gauge group $SO(10)$.
It is natural to imagine a unification within a larger
pseudo-orthogonal group - either $SO(1,13)$ or $SO(3,11)$
or, more exotically $SO(7,7)$.
We shall call such a unified theory a ``graviGUT''
and its main order parameter is now an extended soldering form
$\theta^a{}_\mu$, with $a=1\ldots 14$.
We will not review this in detail (see \cite{Krasnov:2017epi} for a recent review of unified theories) except to recall the following
points.
The fermions of the SM are at the same time spinors
of the Lorentz group and of the GUT group $SO(10)$.
Such fields can be viewed as forming an irreducible representation 
(a 64-real-dimensional Majorana-Weyl spinor) of the graviGUT group $SO(3,11)$ or $SO(7,7)$.
The group $SO(1,13)$ has Weyl representation, but no Majorana-Weyl,
and the decomposition of a 64-complex-dimensional Weyl
gives both a family and an antifamily, that is hard to get rid of.
It is possible to write an action for the fermions,
coupled to the graviGUT gauge field, that reduces to the
correct action of $SO(10)$ fermions coupled to gravity
if one postulates the existence of suitable VEVs.
Natural generalizations of the torsion squared terms,
in the presence of a VEV for the extended soldering form,
give a mass to all components of the graviGUT gauge field
except for $SO(10)$.
The breaking of $SO(10)$ to the SM gauge group, 
requires separate order parameters.

Altogether, one can say that some ingredients of a unified theory are present
and work as expected, but several aspects remain
to be properly understood.

\end{document}